\documentclass[aps,prb,twocolumn,showpacs,superscriptaddress]{revtex4-1}
\usepackage{graphicx}

\begin{document}

\title{Structural evolution and electronic properties of (Sr$_{1-x}$Ca$_x$)$_{2-y}$IrO$_{4+z}$ spin-orbit assisted insulators }

\author{Xiang Chen}
\affiliation{Department of Physics, Boston College, Chestnut Hill, Massachusetts 02467, USA.}
\affiliation{Materials Department, University of California, Santa Barbara, California 93106, USA.}
\author{Stephen D. Wilson}
\email{stephendwilson@engineering.ucsb.edu}
\affiliation{Materials Department, University of California, Santa Barbara, California 93106, USA.}

\date{\today}

\begin{abstract}
The effect of isoelectronic substitution within single crystals of (Sr$_{1-x}$Ca$_x$)$_{2-y}$IrO$_{4+z}$ is explored.  The nominal $n=1$ Ruddlesden-Popper phase with $y=0$, $z=0$ remains stable from $x=0$ until $x=0.11$, where the antiferromagnet spin-orbit Mott insulating state persists. An increase in the saturated moment is observed with increasing Ca-substitution, suggesting a modified coupling of the in-plane moments relative to the in-plane rotation of IrO$_6$ octahedra. Beyond $x=0.11$, the $x=1/4$, $y=0$, $z=1/2$ structural phase Sr$_3$CaIr$_2$O$_9$, consisting of a three-dimensional network of corner sharing octahedra, begins to intermix and eventually nucleates phase pure crystals at higher starting Ca-content.  An insulating, nonmagnetic ground state is observed in this phase attributable to the $J=0$ state and is consistent with a recent powder study. At higher Ca-concentrations beyond $x = 0.75$, crystals begin to stabilize in the $y=1/3$, $z=0$ quasi one-dimensional Ca$_{5}$Ir$_3$O$_{12}$ structure. The low temperature transport in this chain-based structure is well described via variable range hopping, and an antiferromagnetic ordering transition appears below T$_N=9$ K.  Our data provide a detailed mapping of the electronic and structural properties accessible as the structural framework of the canonical spin orbit Mott insulator Sr$_2$IrO$_4$ is destabilized via isovalent chemical substitution.
\end{abstract}


\maketitle

\section{Introduction}
Iridium oxides have received substantial attention due to their potential of realizing various exotic ground states, such as Kitaev spin liquids, Weyl semimetals, and spin orbit Mott insulators.  \cite{Witczak-Krempa2014, Rau2016, Balents2010, HwanChun2015, Wan2011, PhysRevLett.102.017205, Kim2008, Kim2009, Moon2008} One of the key ingredients is strong spin-orbit coupling's interplay with other comparable energy scales, such as the remnant on-site Coulomb repulsion $U$ and crystal field splitting $\Delta$. One consequence is the formation of a spin and orbital angular momenta entangled $J_{eff}=\frac{1}{2}$ wave function for Ir$^{4+}$ valence electrons in a cubic crystal field. Further richness is added via the existence of various stable lattice geometries supporting this wave function, which in turn provide a variety of platforms for hosting distinct quantum states. \cite{Kimchi2014} Prominent examples include the honeycomb lattice Na$_2$IrO$_3$, \cite{Singh2010PRB, Shitade2009, HwanChun2015} hyper-Kagome lattice Na$_4$Ir$_3$O$_8$, \cite{Okamoto2007PRL, Lawler2008PRL} pyrochlore geometries A$_2$Ir$_2$O$_7$ (A=Y, Pr, Nd, Sm, ...)\cite{Matsuhira2011JPSJ, Nakatsuji2006, Machida2010, Ma538} and the layered square lattice Ruddlesden-Popper (R.P.) A$_{n+1}$Ir$_{n}$O$_{3n+1}$ (A=Sr, Ca, Ba) iridates.\cite{Kim2008, Kim2009, Moon2008, Crawford1994, Huang1994, SUBRAMANIAN1994645, JunghoKim2012, Okabe2011, Arita2012, Cao2000657, Ca2002Sr327}

Within the $n=1$ R.P. structure type, a spin orbit-assisted Mott (SOM) state was proposed as a framework for understanding the insulating ground state of Sr$_2$IrO$_4$ (Sr-$214$).\cite{Kim2008, Kim2009, Moon2008}  Although this system has long been of interest due to structural analogies drawn between it and its $3d$-electron high T$_c$ analogue La$_2$CuO$_4$,\cite{Crawford1994, Huang1994, GCao1998} more recent electronic analogies have mapped the single-band Hubbard model of hole doped cuprates into electron doped Sr-$214$ and proposed that unconventional superconductivity may emerge from the carrier doped SOM state.\cite{Wang2011, Watanabe2012, Meng2014} Recent angle-resolved photoemission and scanning tunneling microscopy measurements of surface electron-doped Sr$_2$IrO$_4$ have supported this theoretical mapping by suggesting a gap with $d$-wave symmetry forms at low temperatures.\cite{Kim2015, Yan2015}

Tremendous experimental effort has since focused on perturbing the spin orbit Mott ground state via a variety of methods such as carrier substitution and through applying external pressure,\cite{Korneta2010, MGe2011, TFQi2012, Chen2015, Yklein2008, Lee2012, Cava1994, Yuan2015, Wang2015PRB, Haskel2012, Zocco2014} yet detection of bulk superconductivity remains elusive to date.  This is potentially due to limits imposed by chemical solubility\cite{Chen2015} and extrinsic disorder\cite{Chen2015} such as oxygen vacancies\cite{Korneta2010, Sung2016} and other possible defects; however regardless of these obstacles, various novel phenomena have been uncovered in carrier doped R.P. iridates.  These range from the emergence of nearby competing magnetic states in hole-doped Sr-214, \cite{Calder2012, Calder2015} to electronic phase separation in hole-doped Sr$_3$Ir$_2$O$_7$,\cite{Dhital2014} to negative electronic compressibility in electron-doped Sr$_3$Ir$_2$O$_7$.\cite{He2015} This growing list is a manifestation of the sensitivity of the competing energy scales inherent to the SOM state, and it further points toward the potential of realizing novel electronic and magnetic properties within these compounds as the underlying Mott state is quenched.

\begin{figure*}
\includegraphics[scale=.75]{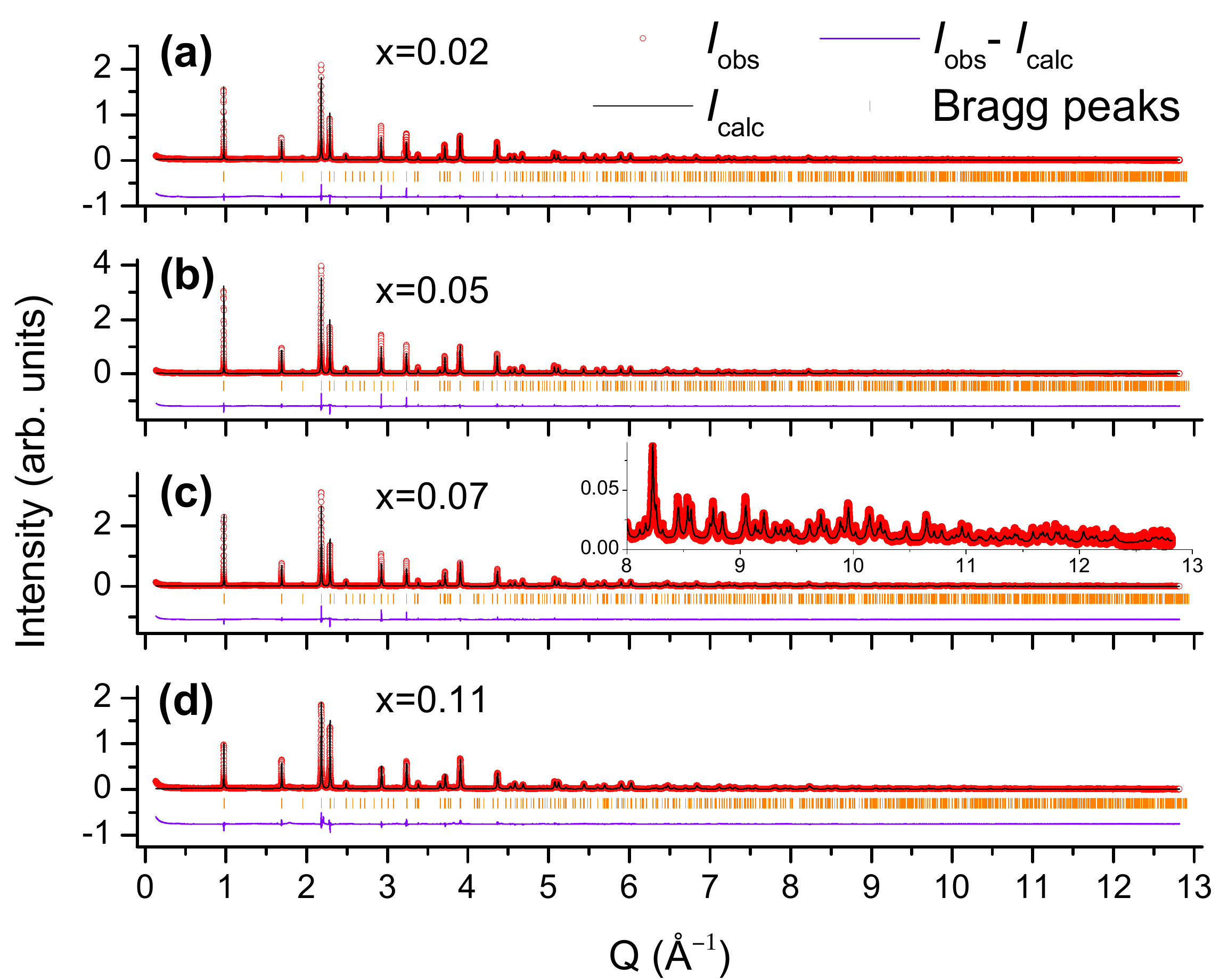}
\caption{Synchrotron powder x-ray diffraction collected at 300 K on select Ca concentrations of (Sr$_{1-x}$Ca$_x$)$_2$IrO$_4$ with (a) $x=0.02$, (b) $x=0.05$, (c) $x=0.07$ (d) $x=0.11$. Inset shows expanded portion of the data and corresponding fit.}
\end{figure*}

Perturbing the bandwidth of SOM compounds via isoelectronic substitution and steric effects presents an alternative approach to destabilizing the SOM state.  As one example, in Sr-214 either Ba or Ca substitution onto the Sr-site (A-site) should decrease or increase the in-plane IrO$_6$ octahedral rotation respectively and ultimately modify the system's bandwidth and noncollinear AF ground state.   Currently however, there are relatively few studies harnessing chemical pressure as a means of modifying the SOM ground states of the RP iridates.  While earlier exploration of polycrystalline samples achieved relatively low levels of A-site substitution,\cite{Shimura1995} more recent single crystal studies have shown that larger substitution levels are still possible.\cite{PhysRevLett.114.257203} For the case of Ca-substitution into Sr-214, the fully substituted end point Ca$_{2-y}$IrO$_4$ is known to form a quasi-one dimensional lattice structure at ambient pressure; however the broader evolution of the structural and electronic phase stability between these two endpoints is an open question.

Here we report a study of isovalent substitution within single crystals of the form (Sr$_{1-x}$Ca$_x$)$_{2-y}$IrO$_{4+z}$, where progressive substitution of Ca$^{2+}$ onto Sr$^{2+}$ sites drives the structure to evolve across a series of structure types, each with insulating ground states where spin-orbit coupling plays an important role. Three distinct structural phases form as a function of increasing Ca content in (Sr$_{1-x}$Ca$_x$)$_{2-y}$IrO$_{4+z}$: For low Ca substitution levels $0 \le x \le 0.11$, the $y=0$, $z=0$ single layer R.P. structure of Sr$_2$IrO$_4$ remains stable. Beyond $x\approx0.11$ a miscibility gap opens and leads to the nucleation of the $x=1/4$, $z=1/2$ phase of Sr$_3$CaIr$_2$O$_9$---a three-dimensional buckled honeycomb lattice of IrO$_6$ octahedra.  For $x>0.7$, the $y=1/3$, $z=0$ quasi one-dimensional chain structure of Ca$_{5}$Ir$_3$O$_{12}$ then stabilizes. Across this phase diagram, charge transport measurements reveal that these three structure types retain their insulating ground states, and resistivity data reflects variable range hopping mechanisms of differing effective dimensionality within each structure-type. The canted antiferromagnetic (AF) structure of Sr-214 persists in lightly substituted (Sr$_{1-x}$Ca$_x$)$_2$IrO$_4$ phase ($0 \le x \le 0.11$) with an enhancement in the saturated ferromagnetic moment observed, attributable to the increase in the $ab$-plane canting of IrO$_6$ octahedra. The Sr$_3$CaIr$_2$O$_9$ phase, on the other hand, remains non-magnetic due to its $J_{eff}=0$ ground state, consistent with a recent powder study.\cite{Wallace2015} The heavily substituted single crystals in the Ca$_{5}$Ir$_3$O$_{12}$ phase show an AF phase transition at T$_N=9$K; however at higher temperatures an anomalous anisotropy is observed in the local moment behavior.  Our data provide a global view of the evolution of the ambient pressure electronic ground states and lattice stabilities of Sr-214 as increasing chemical pressure is introduced via isovalent substitution of smaller Ca cations.

\begin{table*}
\caption {\label{tab:Sr214} Structural and refinement parameters for (Sr$_{1-x}$Ca$_x$)$_2$IrO$_4$ phase ($0 \le x \le 0.11$) at T$=295$K  with space group $I4_1/acd$. The atomic positions of (Sr/Ca), Ir, O$1$ and O$2$ are $(0, 0.25, z)$, $(0, 0.25, 0.375)$, $(x, x+0.25, 0.125)$, $(0, 0.25, z)$ respectively. The isotropic thermal parameters U of O$1$ and O$2$ at $x=0.11$ are constrained to be the same.}
\begin{ruledtabular}
\begin{tabular}{l|lllll}
  (Sr$_{1-x}$Ca$_x$)$_2$IrO$_4$ &$x=0$ &$x=0.02$ &$x=0.05$ &$x=0.07$ &$x=0.11$\\
 \hline
   a(\AA)      &5.49379(1)   &5.49175(2)    &5.49014(1)    &5.48711(2)    &5.48332(4)\\
   c(\AA)      &25.80154(5)  &25.79193(15)  &25.78151(12)  &25.76421(14)  &25.74093(24)  \\
   V(\AA$^3$)   &778.734(2)  &777.866(6)    &777.098(5)    &775.720(5)    &773.948(10) \\
   \\
   Atomic positions and U: \\
   Sr/Ca \\
   z             &0.05051(1)  &0.05033(2) &0.05030(1) &0.05018(1) &0.04996(2) \\
   U(\AA$^2$)    &0.350(7)    &0.282(15)  &0.233(11)  &0.210(12)  &0.285(14)\\
   O$1$ \\
   x             &0.20401(37)  &0.20298(68) &0.20282(50) &0.20255(55) &0.20142(69) \\
   U(\AA$^2$)    &0.402(61)    &0.948(131)  &0.873(95)   &1.026(107)  &2.172(92) \\
   O$2$ \\
   z             &0.95535(10) &0.95631(21) &0.95624(16) &0.95688(18) &0.95664(25) \\
   U(\AA$^2$)    &0.493(45)   &1.034(87)   &1.023(66)   &1.319(73)   &2.172(92) \\

   \\
   Bond length (\AA): \\
   Ir-O$1$    &1.975(3)  &1.976(4)   &1.975(3)  &1.975(4)   &1.975(4) \\
   Ir-O$2$    &2.073(3)  &2.097(6)   &2.094(5)  &2.110(5)   &2.100(8) \\

   Bond angles ($^{\circ}$): \\
   Ir-O$1$-Ir  &159.15(16) &158.7(3)   &158.62(18)  &158.5(2)   &158.0(3) \\

   \\
   \hline
   R-factors: \\
   $\chi^2$  &4.28 &3.64 &4.55 &4.27 &5.43 \\
   $R_{wp}$  &11.9 &13.6 &11.8 &12.1 &12.9 \\
   $R_p$     &8.87 &11.3 &9.47 &9.74 &9.81 \\
\end{tabular}
\end{ruledtabular}
\end{table*}

\section{Experimental Methods}

Single crystals were grown via a platinum (Pt) crucible-based flux growth method similar to procedures reported earlier.\cite{Chen2015} Stoichiometric amounts of precursor materials of SrCO$_3$ ($99.99\%$, Alfa Aesar), CaCO$_3$ ($99.99\%$, Alfa Aesar), IrO$_2$ ($99.99\%$, Alfa Aesar), and anhydrous SrCl$_2$ ($99.5\%$, Alfa Aesar) were weighted in a $2(1-x)$ : $2x$: $1$ : $6$ molar ratio, where $x$ is the nominal Ca concentration. The starting powders were fully ground, mixed and sealed inside a Pt crucible, covered by a Pt lid, and placed inside an outer alumina crucible. Mixtures were heated slowly to $1380^\circ$C, soaked for $5$ hours, slowly cooled to $850^\circ$C over $120$ hours and then furnace cooled to room temperature over $\approx5$ hours. Single crystals were then obtained after dissolving excess flux with deionized water.

As an initial check of the structural phases of crystals nucleated, single crystals within a single batch were randomly chosen, ground into powder, and measured in a PANalytical Empyrean x-ray diffractometer at room temperature. Both phase pure and mixed phase batches were found, depending the nominal doping concentration $x$.  For mixed phase batches, the relative percentage of each constituent phase was obtained from the Rietveld refinement of powder x-ray data with the Fullprof Suite.\cite{RodríguezCarvajal199355} Elemental concentrations and atomic ratios were also determined via energy dispersive spectroscopy (EDS) measurements. Typical standard deviations of the relative Ca concentrations found within one growth batch were less than $1\%$. Measured standard deviations of Ca content are shown as horizontal error bars in plots versus Ca-concentration and were obtained by repeated EDS measurements on random spots across a crystal.

Only single crystals grown from batches with starting Ca concentrations that yielded phase pure crystals were chosen for further detailed characterization and measurement.  Select crystals were crushed into powders and measured on the $11$-BM beam line at the Advanced Photon Source at Argonne National Laboratory for synchrotron powder X-ray diffraction (PXRD) measurements.  Transport measurements were performed within a Quantum Design Physical Property Measurement System (PPMS), and magnetization measurements were performed using a Quantum Design MPMS $5$XL SQUID magnetometer.

\section{Experimental Results}
\subsubsection{$0 \le x \le 0.11$, $y=0$, $z=0$, (Sr$_{1-x}$Ca$_x$)$_2$IrO$_4$ Ruddlesden-Popper phase }

\begin{figure}
\includegraphics[scale=.34]{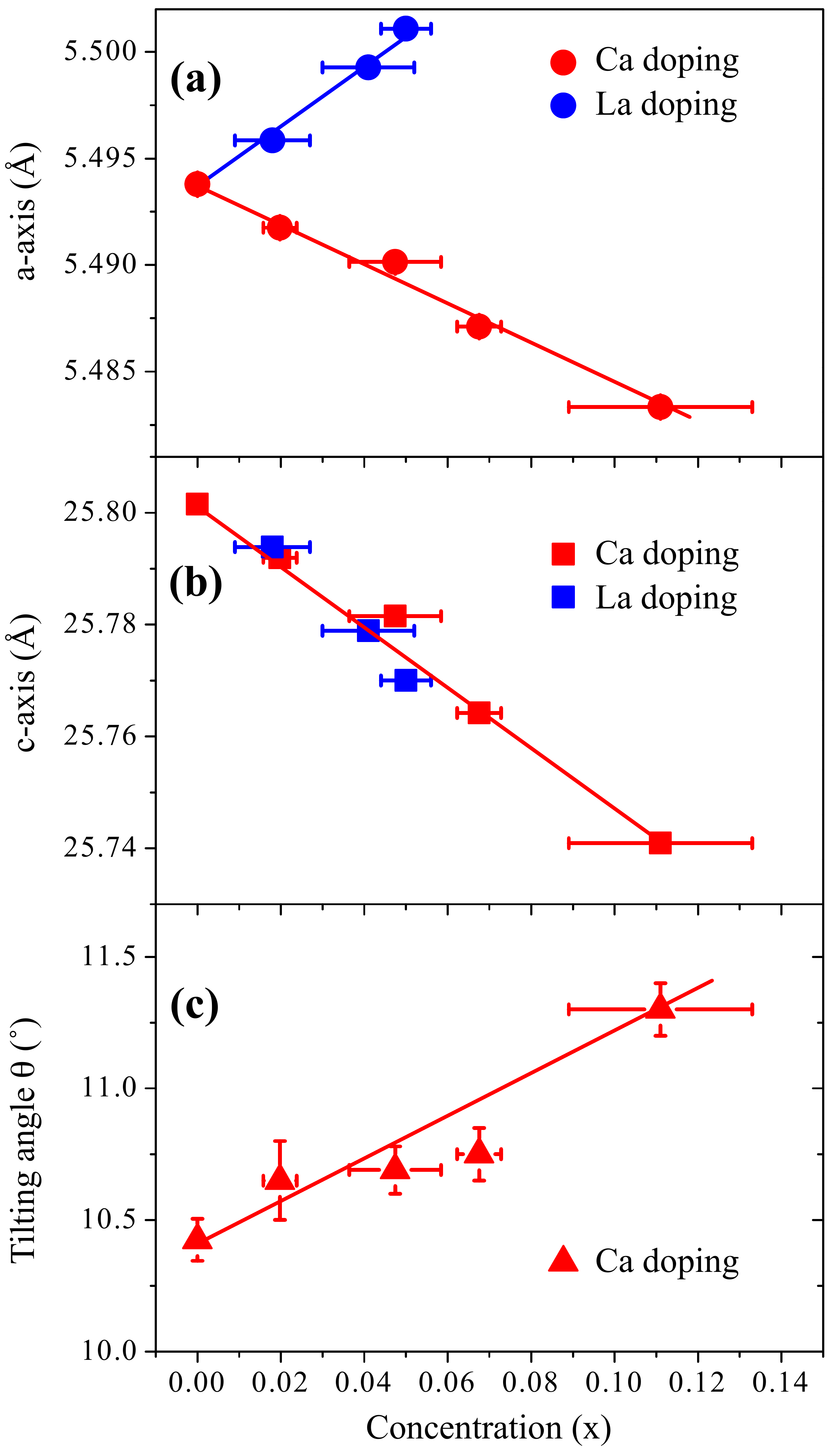}
\caption{Lattice parameters and in-plane IrO$_6$ canting angles $\theta$ determined at 300 K from synchrotron PXRD refinement. (a) a-axis lattice parameters, (b) c-axis lattice parameters, and (c) in-plane IrO$_6$ octahedral canting angles $\theta$ are plotted as a function of Ca-substitution $x$.  For comparison, data resulting from La$^{3+}$ substitution reported previously \cite{Chen2015} are overplotted with effect of Ca$^{2+}$ substitution. Red symbols represent Ca substitution, while blue symbols represent La doping.}
\end{figure}

For starting compositions of $0 \le x_{nominal} \le 0.12$, the $n=1$ R.P. structure remained stable, and batches yielded phase pure crystals of (Sr$_{1-x}$Ca$_x$)$_2$IrO$_4$. Select crystals in this doping regime were crushed and synchrotron powder XRD data collected at $300$ K.  As shown in Fig $1$., the data can be indexed and fit within the space group $I4_1/acd$; however we note that the distinction between this and the lower symmetry space group $I4_1/a$ recently reported for parent Sr$_2$IrO$_4$ system \cite{Torchinsky2015, Ye2015} is beyond the resolution of our measurements. No further symmetry lowering is observed upon Ca substitution for $x<0.10$ and no additional structural phases were observed within the ground crystals.  Crystals with Ca-concentrations from a nominal $x_{nominal}=0.15$ batch began to show inclusions of intergrowths of the Sr$_3$CaIr$_2$O$_9$ structure (discussed further in section III.2) at a level of $\approx4\%$ as well as a small inclusion of the $n=2$ R.P. phase at a level of $\approx6\%$.

\begin{figure}
\includegraphics[scale=.425]{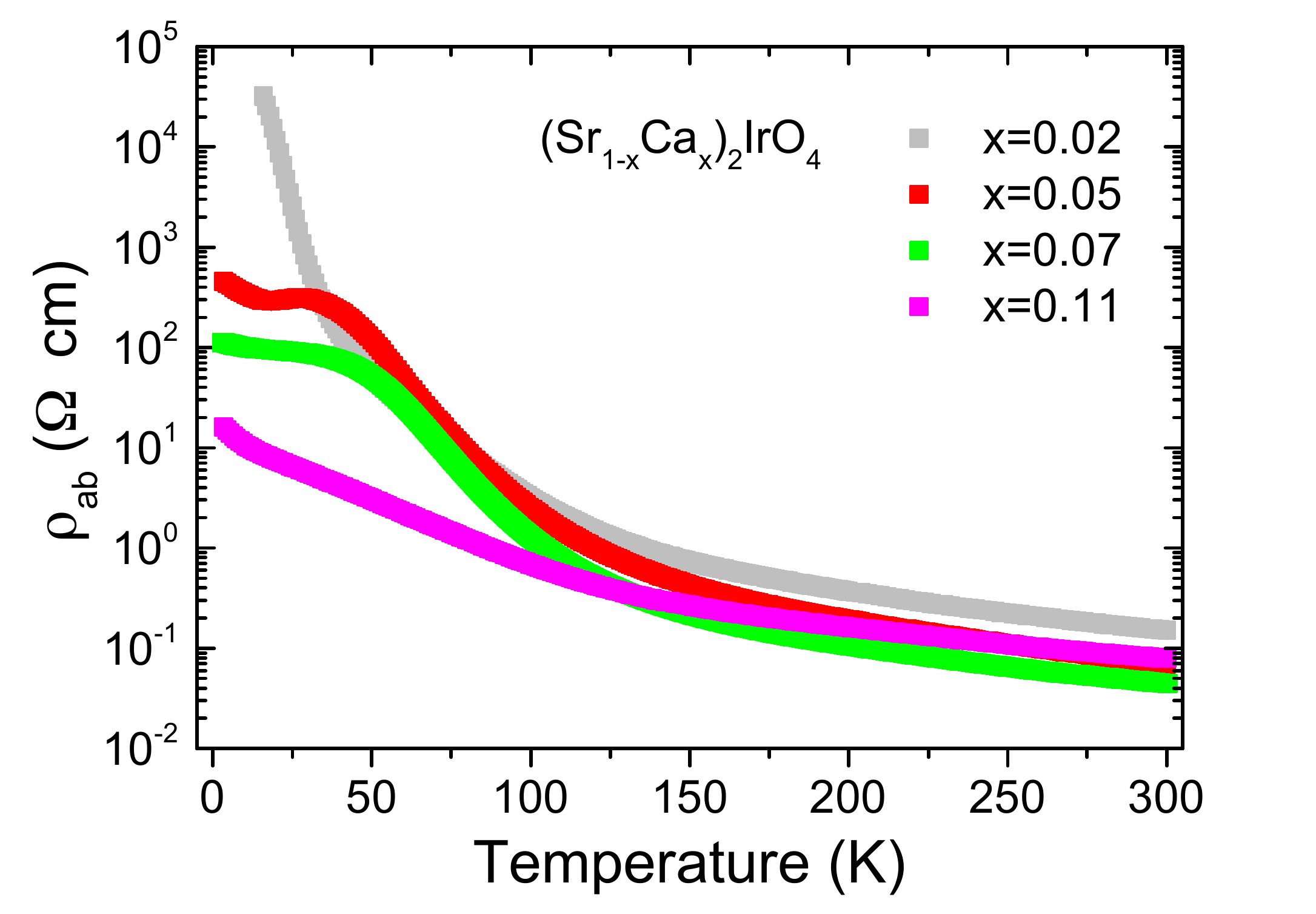}
\caption{The temperature dependence of the $ab$-plane resistivity of (Sr$_{1-x}$Ca$_x$)$_2$IrO$_4$ plotted for Ca concentrations of $x=0.02$ (grey symbols), $x=0.05$ (red symbols), $x=0.07$ (green symbols), and $x=0.11$ (pink symbols).}
\end{figure}

The structural parameters obtained from refinement of crushed single crystal powder XRD data are summarized in Table I. From these data, the lattice parameters $a$, $c$ and the in-plane octahedral canting angles $\theta$ were extracted and plotted in Figure 2, where $\theta=(180^{\circ} - \alpha)/2$ and $\alpha$ is the Ir-O-Ir nearest neighbor bond angle. For comparison, corresponding data for La doping (blue symbols) are also included. As Sr$^{2+}$ is replaced by the smaller Ca$^{2+}$ ion,\cite{Shimura1995} both the lattice $a$ and $c$ axes decrease as expected. This contrasts the case of when Sr$^{2+}$ is replaced by the smaller electron donor La$^{3+}$ and preferentially swells the $a$ axis, \cite{Chen2015} likely by adding carriers into the antibonding $e_g$ orbitals.

The low temperature insulating ground state persists through the limit of Ca solubility ($x\approx0.11$) in the $n=1$ R.P. Sr-$214$ phase. Charge transport data for select Ca concentrations are shown in Fig. $3$ where increased levels of Ca impurities reduce the low temperature resistivity as this limit is approached. Resistivity data did not fit to a thermally activated hard gap or variable range hopping (VRH) models over an appreciable temperature range. We note that the $x=0$ parent material is known to display three-dimensional VRH at low temperatures; \cite{Kini2006} however this is likely obscured by the added effect of Ca-impurities.  The low temperature plateau in resistivity apparent in $x=0.05$ and $x=0.07$ samples seemingly suggests the approach of a percolative metal-insulator transition; however no insulator to metal transition was observed below $300$ K in any of the measured crystals.

\begin{figure}
\includegraphics[scale=.425]{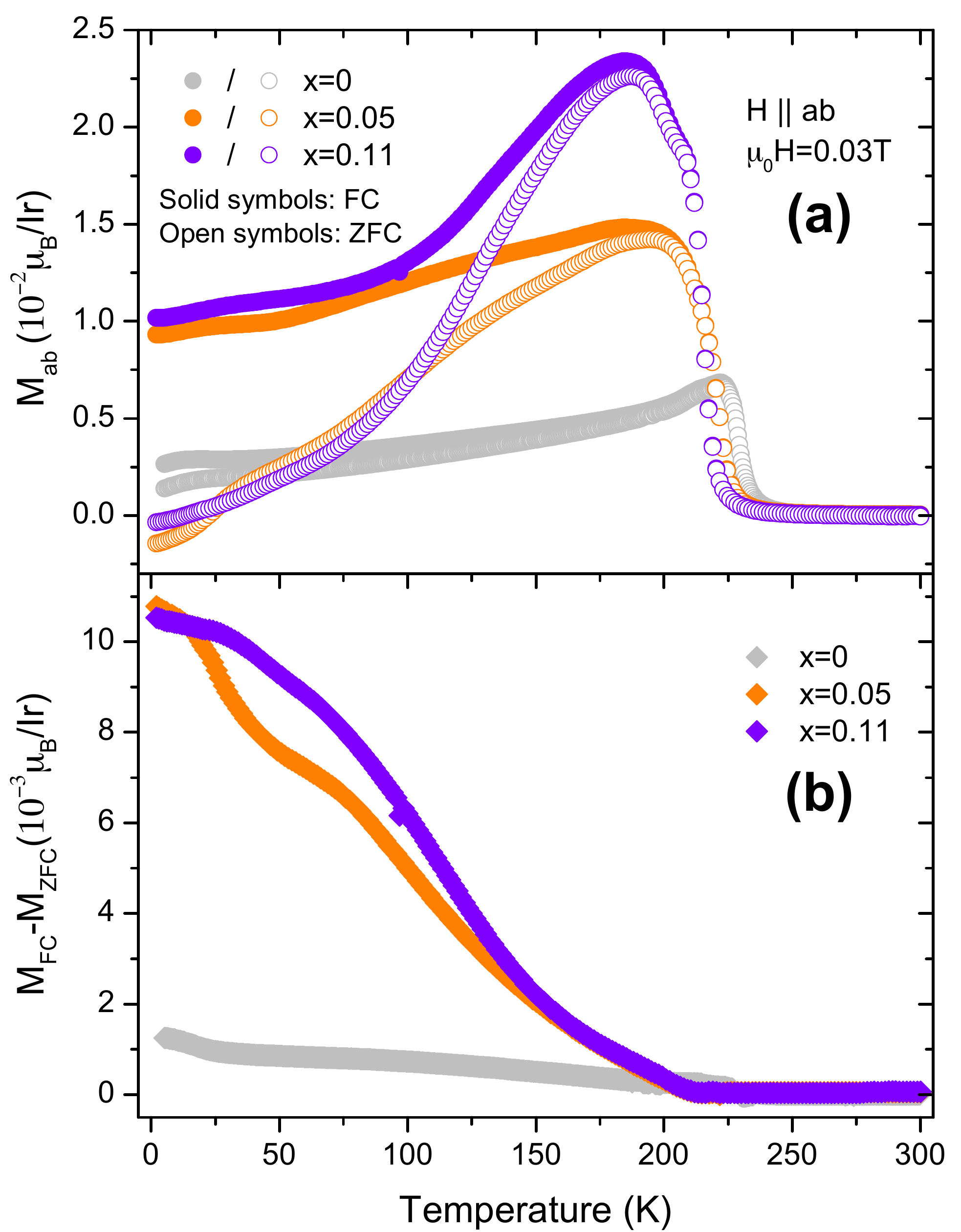}
\caption{ Magnetization data collected on (Sr$_{1-x}$Ca$_x$)$_2$IrO$_4$ samples. (a) Temperature dependent magnetization data collected under ZFC and FC with the magnetic field applied in the $ab$-plane and $\mu_0H=0.03$ T. (b) Irreversibility of the magnetization data plotted in panel (a) determined by subtracting FC and ZFC data.}
\end{figure}

In order to further parameterize the evolution of the SOM state in Sr-214 upon Ca-substitution, magnetization measurements were performed.  Figure 4 shows field cooled (FC) and zero-field cooled (ZFC) magnetization data collected under an $\mu_0H=300$ Oe applied within the $ab$-plane (Fig. $4$(a)). Upon cooling under a fixed field, the maximum in susceptibility associated with the onset of AF order as well as the magnitude of irreversibility associated with the AF transition are enhanced with progressive Ca substitution.  The enhancement in irreversibility is better illustrated in Fig. 4 (b) where the introduction of Ca impurities is shown to reduce the onset of irreversibility and enhance the low field polarization of the canted spin structure.  The latter point is further illustrated  in Fig. 5 (a) where $T=5$ K isothermal magnetization curves are plotted.  Namely, from the inset of Fig. 5 (a), there is an enhanced remnance of the $ab$-plane magnetization of the canted AF spin structure.  This likely reflects the role of disorder in Ca-substituted samples where Ca-substitution within the rock-salt layers disrupts the extended spin canting sequence between layers and prevents the spin structure from relaxing when the field is removed.  As a result, after cycling the field the sample retains a net ferromagnetic moment even under zero applied field.

\begin{figure}
\includegraphics[scale=.45]{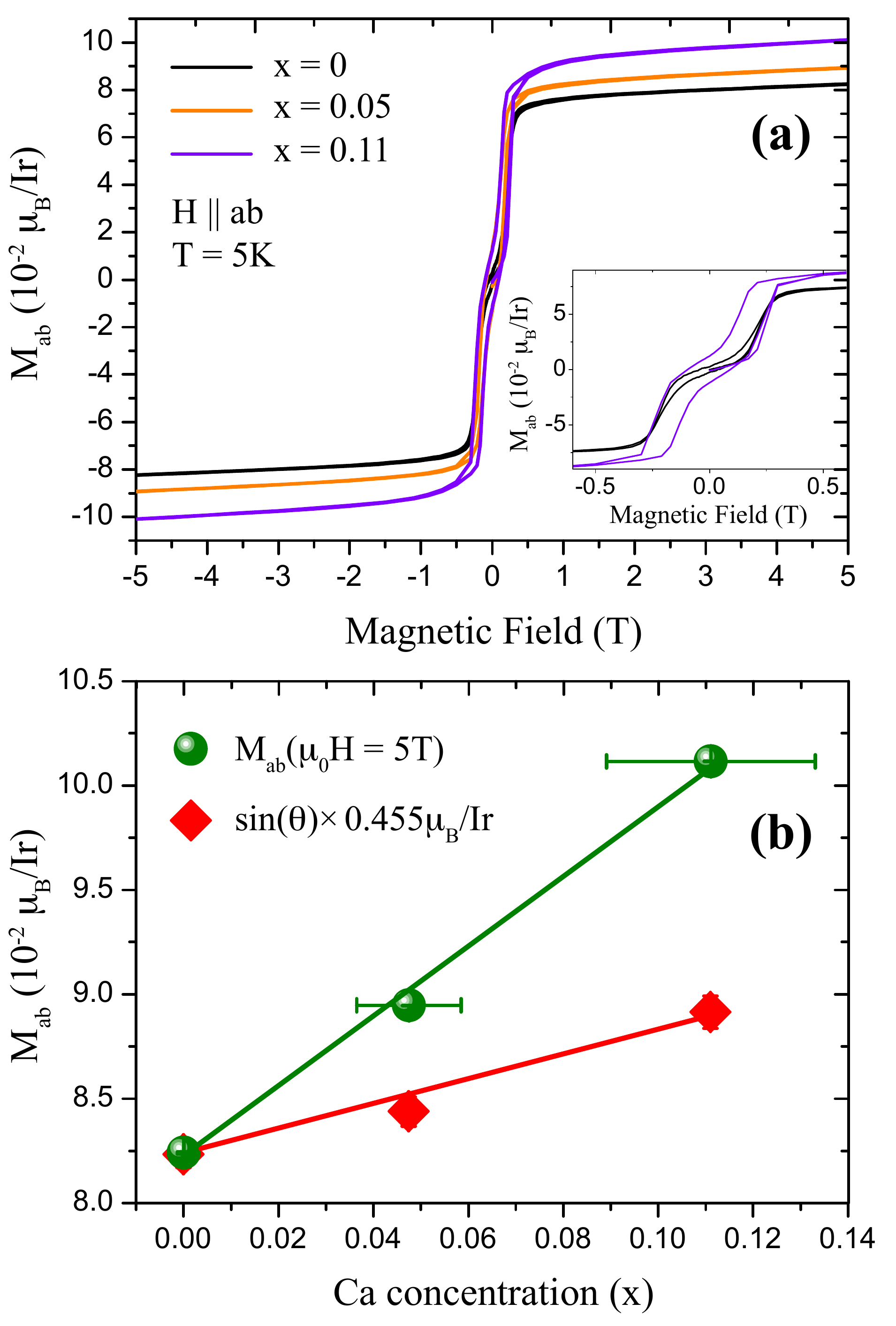}
\caption{(a) $ab$-plane isothermal magnetization of (Sr$_{1-x}$Ca$_x$)$_2$IrO$_4$ collected at T$=5$K. Inset shows an expanded subset of the data with the field swept between $-0.6$T to $0.6$T. (b) Saturated magnetization plotted as a function of Ca concentration $x$ and collected with $\mu_0H=5$T applied parallel to the $ab$-plane.  Overplotted are the canted projections of the net magnetization using an assumed ordered moment $m_{AF}=0.455$ $\mu_B$/Ir and the measured IrO$_6$ octahedral canting angles $\theta$.}
\end{figure}

The field-driven transition out of the zero net moment AF state and into the spin phase with aligned moment canting persists for all Ca concentrations at nearly the same critical field of $H_c\approx0.2$T.\cite{Kim2009} The remnant spin reorientation implies that the magnetic structure remains qualitatively similar to the unalloyed Sr-214 system \cite{Dhital2012, Ye2013} despite the slight suppression of the onset of AF order in Ca-substituted samples from T$_{AF}= 240$K for $x=0$  to T$_{AF}\approx 210$K for $x=0.11$.  An increase in the coercivity of the field-induced transition occurs with increasing Ca-content and is consistent with a disorder induced pinning of the domain structure.  Additionally, the magnitude of the saturated magnetization in the aligned, canted state increases with increasing Ca-content.

Curiously, the increase in the saturated magnetization cannot be explained by a trivial geometric factor such as moments locked to an increase in the $ab$-plane IrO$_6$ octahedral rotation.\cite{0953-8984-25-42-422202}  As a simple estimation, by assuming a fixed moment of $\approx 0.46 \mu_B$ on the Ir-site (close to the zero field value of $0.36 \pm 0.06 \mu_B$ \cite{Dhital2012}) and using the refined canting angle of $10.5^{\circ}$, the canted Ir moments within each layer should give rise to the net magnetization of $0.083$ $\mu_B/Ir$ observed under $5$ T (assuming a single polarized magnetic domain).   If the ordered moment remains fixed as Ca is substituted, the net moment should increase as sin($\theta$), where $\theta$ can be obtained from the PXRD structural refinement.  However, the refined in-plane Ir-O-Ir angles decrease only slightly ($<1\%$) in the most heavily Ca-substituted samples relative to the $x=0$ parent system, while the saturation magnetization increases more than than $23\%$.  This discrepancy between the field polarized net ferromagnetic moments and the expectations of fixed Ir moments locked to the octahedral rotations is illustrated in Fig. 5 (b).  A further enhancement in the magnitude of the ordered moment suggests that a modified canting of the Ir-moments relative to the in-plane IrO$_6$ octahedral rotation should occur; however a future full neutron refinement of both the oxygen positions and ordered moment is necessary to fully explore this possibility.

\subsubsection{$x=1/4$, $y=0$, $z=1/2$, Sr$_3$CaIr$_2$O$_9$ honeycomb lattice phase}

When the starting Ca-concentration was increased to $0.15 \le x_{nominal} \le 0.70$, cube-like single crystals with typical dimensions of $0.9\times0.9\times0.8$ $mm^3$  were obtained. Synchrotron PXRD (Fig.$7$) data were collected on crushed single crystals with this morphology, and the data agree well with the recently reported Sr$_3$CaIr$_2$O$_9$ structure type.\cite{Wallace2015}  EDS measurements on this phase indicated atomic ratios of (Sr+Ca):Ir as 2:1 and Ca:(Ca+Sr) as 1:3.7 .  The crystal structure is shown in Fig $6$ where all of the octahedra (Ir and Ca octahedra) are corner sharing and form a three-dimensional network. The Ir cations sit on two chemically distinct sites and are labelled as Ir$1$ and Ir$2$ respectively. Looking down the $b$ axis, the Ir octahedra are connected through a corner sharing network and form buckled planes, separated by layers of Ca octahedra. Looking down the a$^*$ direction (perpendicular to the $bc$ plane) the Ir octahedra form a buckled honeycomb lattice as shown from the projected view in Fig $6$(b). We note here that Ir cations in this system are nominally in the d$^4$ nonmagnetic $J=0$ state.\cite{Wallace2015} PXRD data were collected at three temperatures ($T =100$, $295$, and $473$ K) and fit within the space group $P2_1/c$. No signature of a structural phase transition was observed, and structural parameters are summarized in Table II. Due to the nine unique oxygen sites and their small contribution to the PXRD profiles, refinement of the oxygen positions and atomic displacement parameters were constrained.

\begin{figure}
\includegraphics[scale=.30]{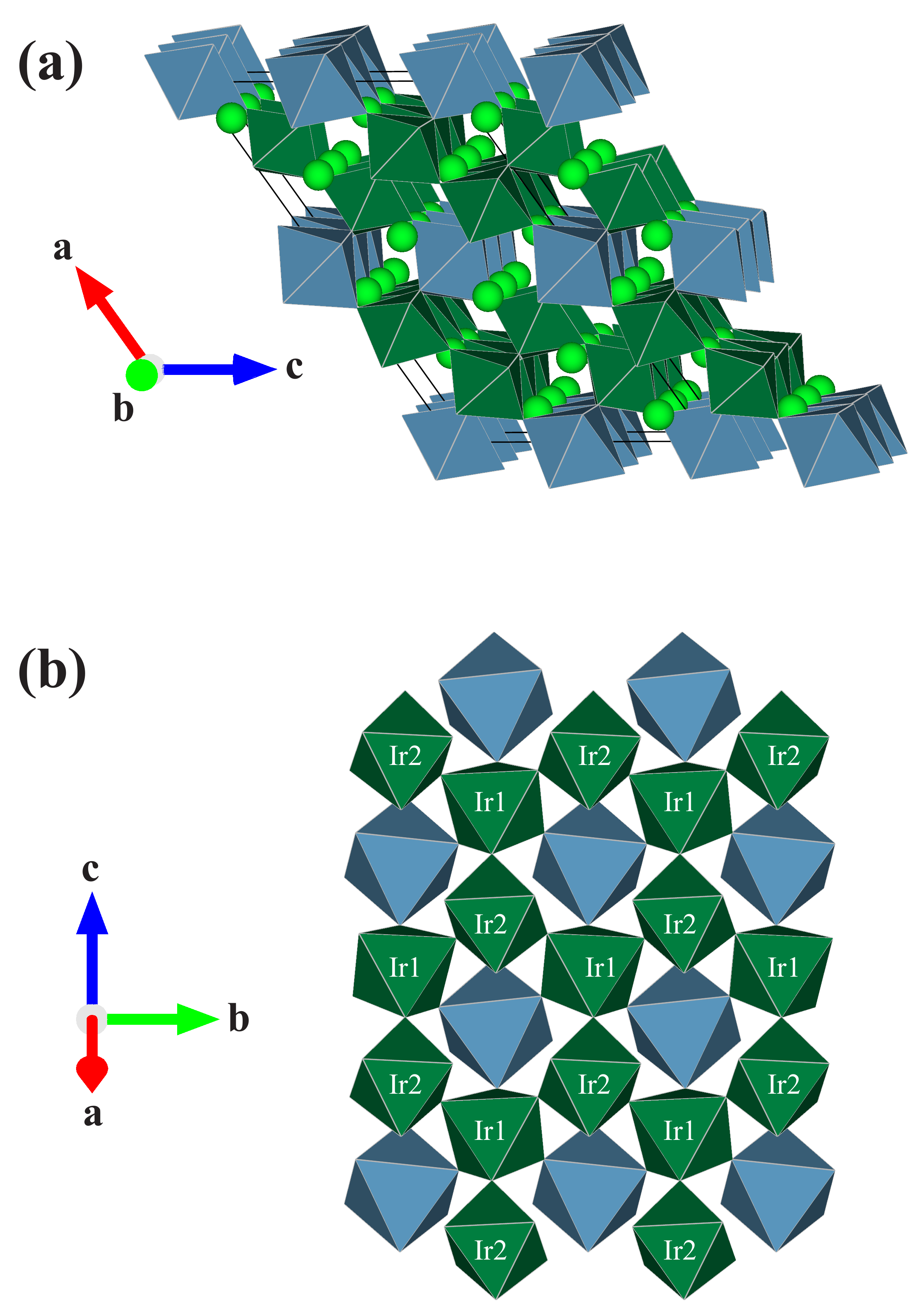}
\caption{(a) Crystal structure of Sr$_3$CaIr$_2$O$_9$. For oxygen at the vertices of Ca and Ir centered octahedra are omitted. Sr atoms are represented by green balls. Ca-centered octahedra are plotted in blue and Ir-centered octahedra are plotted in green.  (b) Projected $bc$-plane view of the lowest three layers of Sr$_3$CaIr$_2$O$_9$ from panel (a) showing the formation of buckled honeycomb lattice of Ir-octahedra.  Distinct Ir-octahedra positions are labeled as Ir1 and Ir2.  VESTA software was used for crystal structure visualization.\cite{Momma2011} }
\end{figure}

\begin{figure*}
\includegraphics[scale=.75]{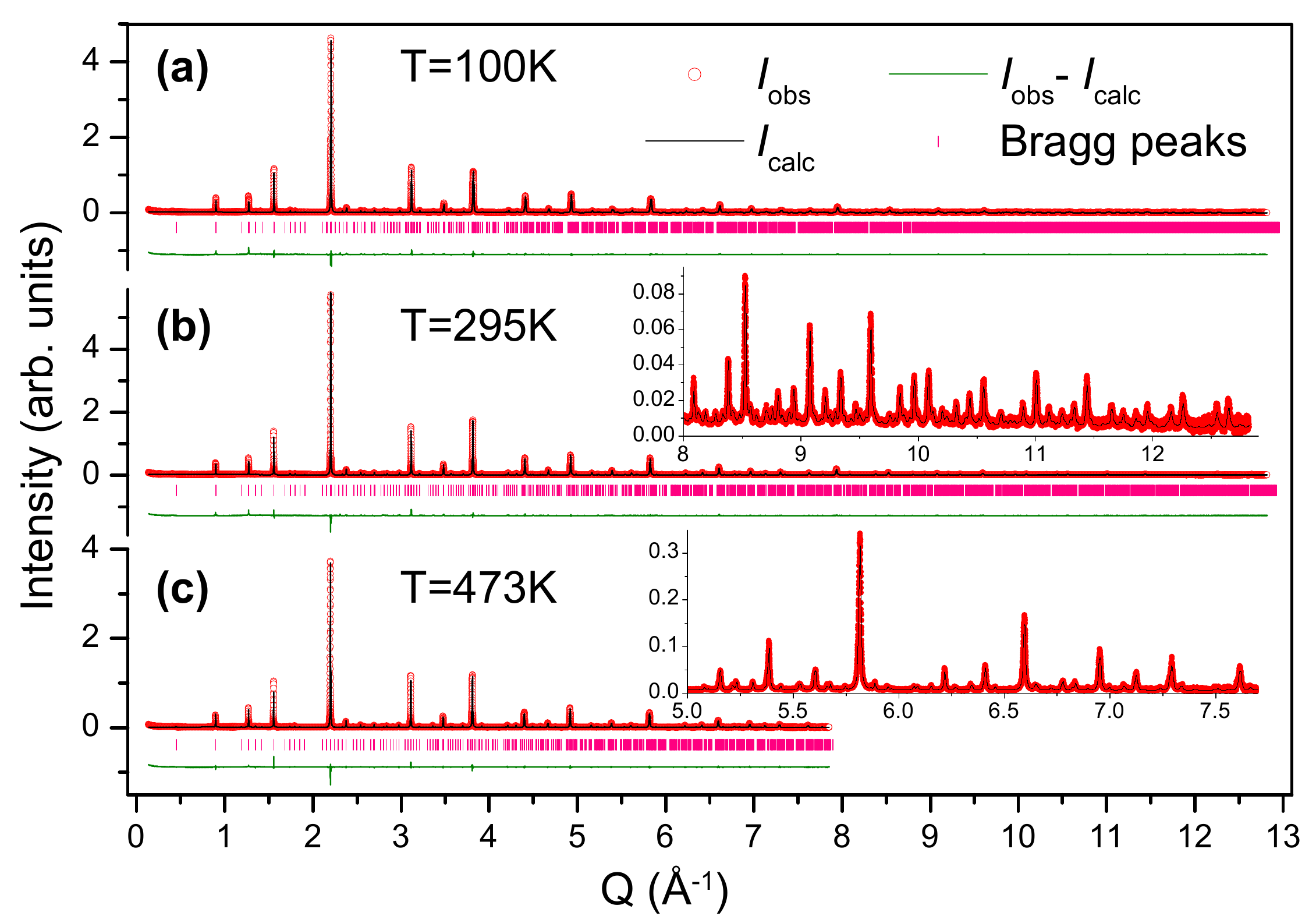}
\caption{ Synchrotron PXRD of crushed Sr$_3$CaIr$_2$O$_9$ single crystals collected at (a) 100 K, (b) 295 K, and (c) 473 K. The data are indexed fit within space group $P2_1/c$. Inset in panel (c) shows the magnified view of the data and fit.}
\end{figure*}

To explore the electronic properties of the Sr$_3$CaIr$_2$O$_9$ phase, charge transport measurements were also performed. Fig. $9$ shows the resistivity data of a single crystal of Sr$_3$CaIr$_2$O$_9$ along the (1,1,1) axis\cite{Supplemental} where the expected spin-orbit stabilized insulating state comprised of a filled $J_{eff}=3/2$ band is observed.  A variable range hopping (VRH) model was applied to model the low temperature resistivity: $\rho=\rho_0\cdot exp((T_0/T)^{1/(1+D)})$, where $D$ is the dimensionality of charge transport and $T_0$ is assumed to be a temperature independent constant. The best fit was achieved using a dimensionality of $D=2$, consistent with an earlier study of powder samples.\cite{Wallace2015}  Illustrating this, Fig. $9$ inset shows ln $\rho$ versus $(\frac{1}{T})^\frac{1}{3}$, where data fit between $2\geq T \geq 125$ K render an effective $T_0\approx 5800K$. Naively, this suggests an underlying conduction mechanism facilitated via tunneling of carriers constrained to planes within the lattice; for instance via an effective layered crystal structure of buckled IrO$_6$ planes located between Ca spacer layers.  Transport measurements along orthogonal crystal axes however failed to resolve any appreciable anisotropy. Absent a detailed understanding of potential twinning in the crystal, more detailed analysis is presently precluded.

\begin{table}[b]
\caption {\label{tab:Sr3129} Structural and refinement parameters for Sr$_{3}$CaIr$_2$O$_9$ phase at select temperatures. The space group $P2_1/c$, initial atom positions and fixed thermal parameters are adopted from D.C. Wallace \textit{et al.}\cite{Wallace2015} }
\begin{ruledtabular}
\begin{tabular}{l|lll}
  Sr$_{3}$CaIr$_2$O$_9$ &$T=100K$ &$T=295K$ &$T=473K$\\
 \hline
   a(\AA)      &17.09503(9)   &17.12849(20)    &17.16315(17)\\
   b(\AA)      &5.71275(3)    &5.71084(5)      &5.71382(7)\\
   c(\AA)      &9.86065(7)    &9.88008(9)      &9.89488(11)  \\
   $\beta$($^{\circ}$)     &125.28085(45)    &125.25529(110)      &125.24737(72)  \\
   V(\AA$^3$)   &786.117(8)  &789.192(13)    &792.465(15)     \\
   \hline
   R-factors: \\
   $\chi^2$  &3.59 &3.76 &4.25  \\
   $R_{wp}$  &11.6 &12.3 &12.8  \\
   $R_p$     &9.18 &9.99 &9.91  \\
\end{tabular}
\end{ruledtabular}
\end{table}

The magnetic properties of Sr$_3$CaIr$_2$O$_9$ were also explored and temperature dependent magnetization data collected under an applied field of $\mu_0H=0.03$ T along the (1,1,1)-axis are plotted in Fig. $10$. Negligible anisotropy was observed upon orienting the $H$-field along orthogonal axes. Rather, the observed magnetization is weak and the measured susceptibility can be captured via a Curie-Weiss (CW) fit of the form $\chi=\chi_0+\frac{C}{T-\Theta_{CW}}$ where $\chi_0$ is the temperature independent constant, $C=\frac{\mu_{eff}^2}{3k_B}$, $\mu_{eff}$ is the effective local magnetic moment, and $\Theta_{CW}$ is the Curie-Weiss temperature. No irreversibility was resolved between the FC and ZFC  magnetization data, the CW temperature $\Theta_{CW}=-1.7(1)$K is near zero, and $\mu_{eff}=0.18(1)$ $\mu_B$.  Taken together this suggests that this weak paramagnetic response has an extrinsic, impurity driven origin.

To further explore this, isothermal magnetization measurements were performed at $T=5$K with the results shown in Fig. 10 (b). No field sweep hysteresis was observed, and for clarity, Fig. 10 (b) shows only the data with the magnetic field sweeping from $0$ to $5$ T. The data can modeled with a simple Brillouin function: $M=\eta g\mu_BJ\cdot(\frac{2J+1}{2J}\coth(\frac{2J+1}{2J}x)-\frac{1}{2J}\coth(\frac{1}{2J}x))$ where $M$ is the magnetization per Ir, $g$ is the $g$-factor, and $x=\frac{g\mu_BJB}{k_BT}$. Assuming an isotropic $g=2$ and $J=\frac{1}{2}$, the data was fit with the impurity fraction scaling factor $\eta=0.015(3)$.  Using this impurity fraction scaling factor, the corrected effective moment $\mu_{eff}$ obtained from the CW fit to data in Fig. 10 (a) becomes $1.53(9)\mu_B$ per impurity site.  This suggests that the observed susceptibility arises from a $2\%$ molar fraction of Ir$^{4+}$ moments in the sample; likely driven via a small number ($\approx 1\%$) of oxygen vacancies.

\begin{figure}
\includegraphics[scale=.375]{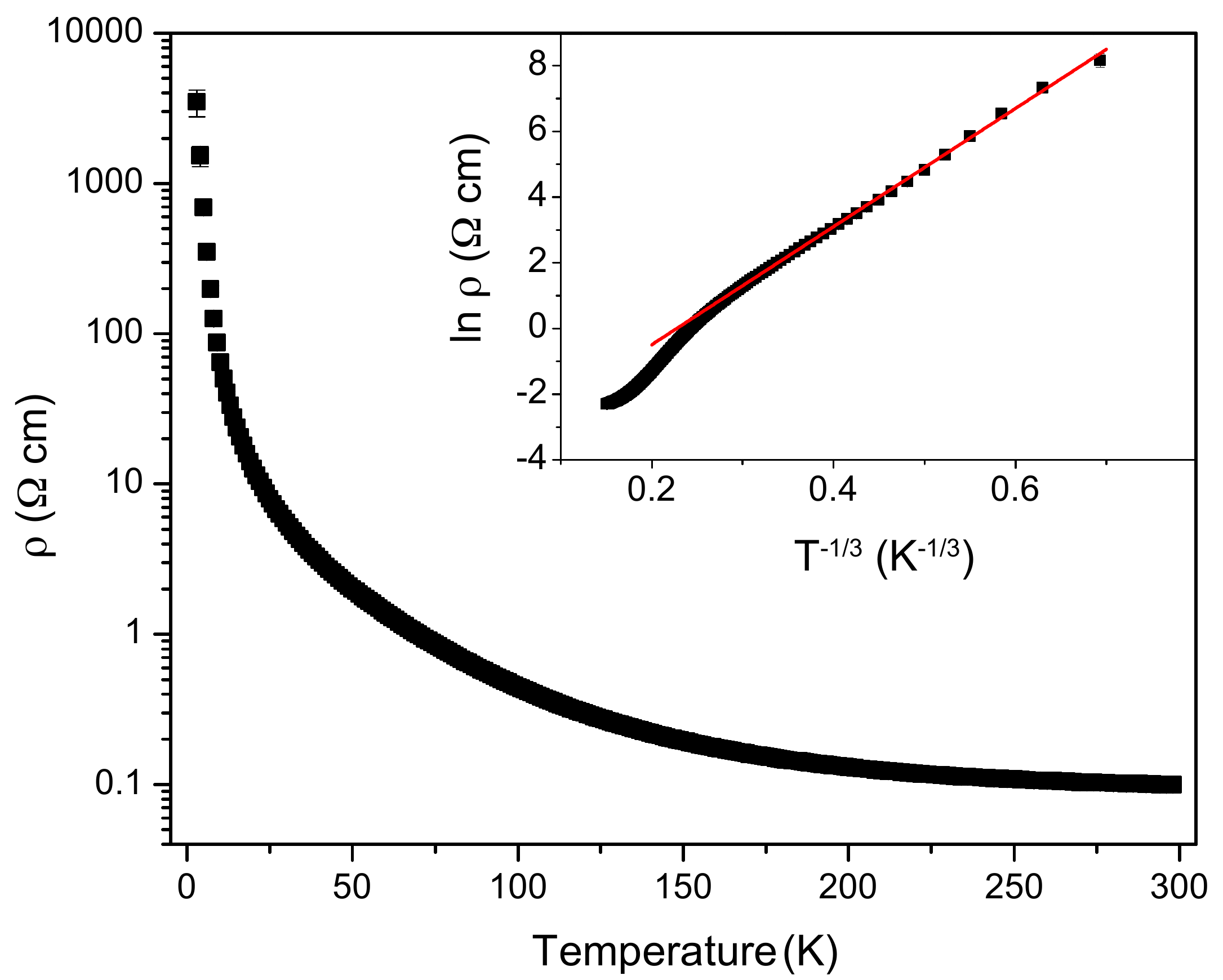}
\caption{Resistivity plotted as a function of temperature measured on a Sr$_3$CaIr$_2$O$_9$ single crystal. Inset shows the same data plotted as $ln\rho$ versus $T^{-\frac{1}{3}}$ where the red line shows a fit to VRH behavior with $D=2$ and T$_0=5800$K.}
\end{figure}

\begin{figure}
\includegraphics[scale=.4]{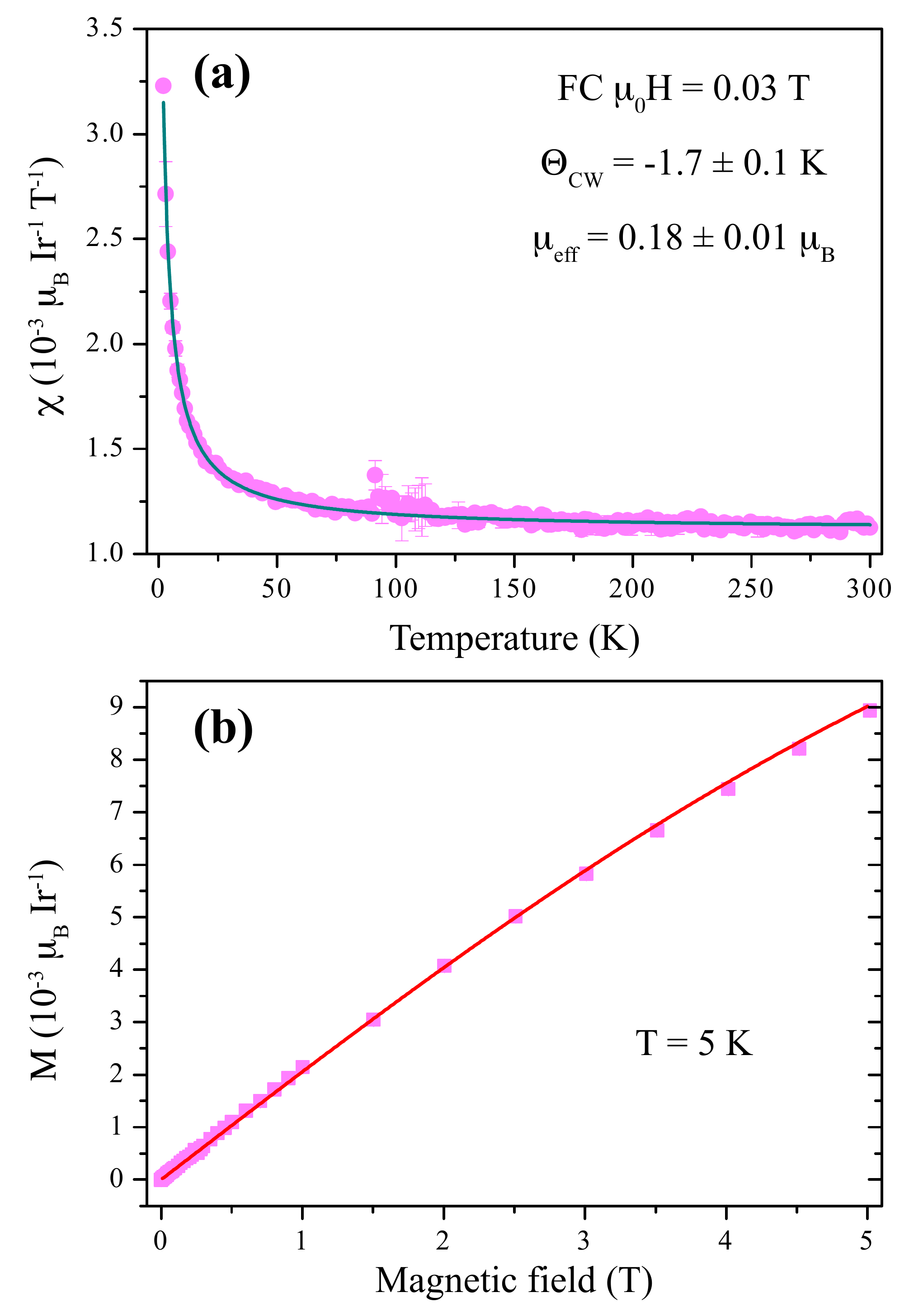}
\caption{(a) Temperature dependent susceptibility of Sr$_3$CaIr$_2$O$_9$ under an applied magnetic field of $0.03$T. Dark cyan line is the Curie-Weiss fit across the whole temperature range, which lead to a small Curie temperature $\Theta_{CW}=-1.7\pm0.1$K and a small effective moment $\mu_{eff}=0.18\pm0.01\mu_B$. (b) Isothermal magnetization measured at T$=5$K. Red line is the Brillouin function fit multiplied by an additional scaling constant $0.015$, assuming $g=2$ and $J=1/2$. The necessity of applying the scaling constant is because not all Ir contribute to the magnetism, whereas the magnetization is normalized to the total number of Ir sites.}
\end{figure}

\subsubsection{$ x\geq0.75$, $y=1/3$, $z=0$, (Sr$_{1-x}$Ca$_{x}$)$_{5}$Ir$_{3}$O$_{12}$ edge-sharing IrO$_6$ chain phase }

When the starting Ca concentration reached $x = 0.75$, crystals possessing the quasi one-dimensional Ca$_{2-y}$IrO$_4$ structure began to nucleate. \cite{Buschbaum200349} It is known that Ca$_2$IrO$_4$ is prone to hosting a substantial amount of Ca vacancies, which results in the formation of the more stable Ca$_5$Ir$_3$O$_{12}$ phase.\cite{Shi2010}  Single crystals were removed, crushed, and PXRD data collected.  PXRD data were best fit within the space group $P\overline{6}2m$ using the previously reported structure of Ca$_5$Ir$_3$O$_{12}$.\cite{Babel1966, Wakeshima2003} EDS measurements revealed the atomic ratio of Ca:(Ca+Sr) to be $0.89(2)$; however, due to strong preferred orientation, reliable (Ca/Sr)-site occupancy information was not obtained within our PXRD analysis.  A small inclusion of the Sr$_3$CaIr$_2$O$_9$ phase also appeared in the PXRD data with a relative molar fraction refined to be less than $0.5\%$.  Lattice parameters were determined to be $a=b=9.45626(14)$\AA, $c=3.20520(6)$\AA\quad, slightly larger than those reported for Ca$_5$Ir$_3$O$_{12}$ ($a=b=9.4208$\AA, $c=3.1941$\AA)\cite{Cao2007PRB, Wakeshima2003} and consistent with the partial substitution of larger Sr$^{2+}$ cations onto Ca$^{2+}$-sites. Given the dearth of evidence supporting the stability of bulk stoichiometric Ca$_{2}$IrO$_4$, we assign the Ca$_5$Ir$_3$O$_{12}$ phase to the new crystals found nucleating in this $x \geq 0.75$ regime.

\begin{figure}
\includegraphics[scale=.375]{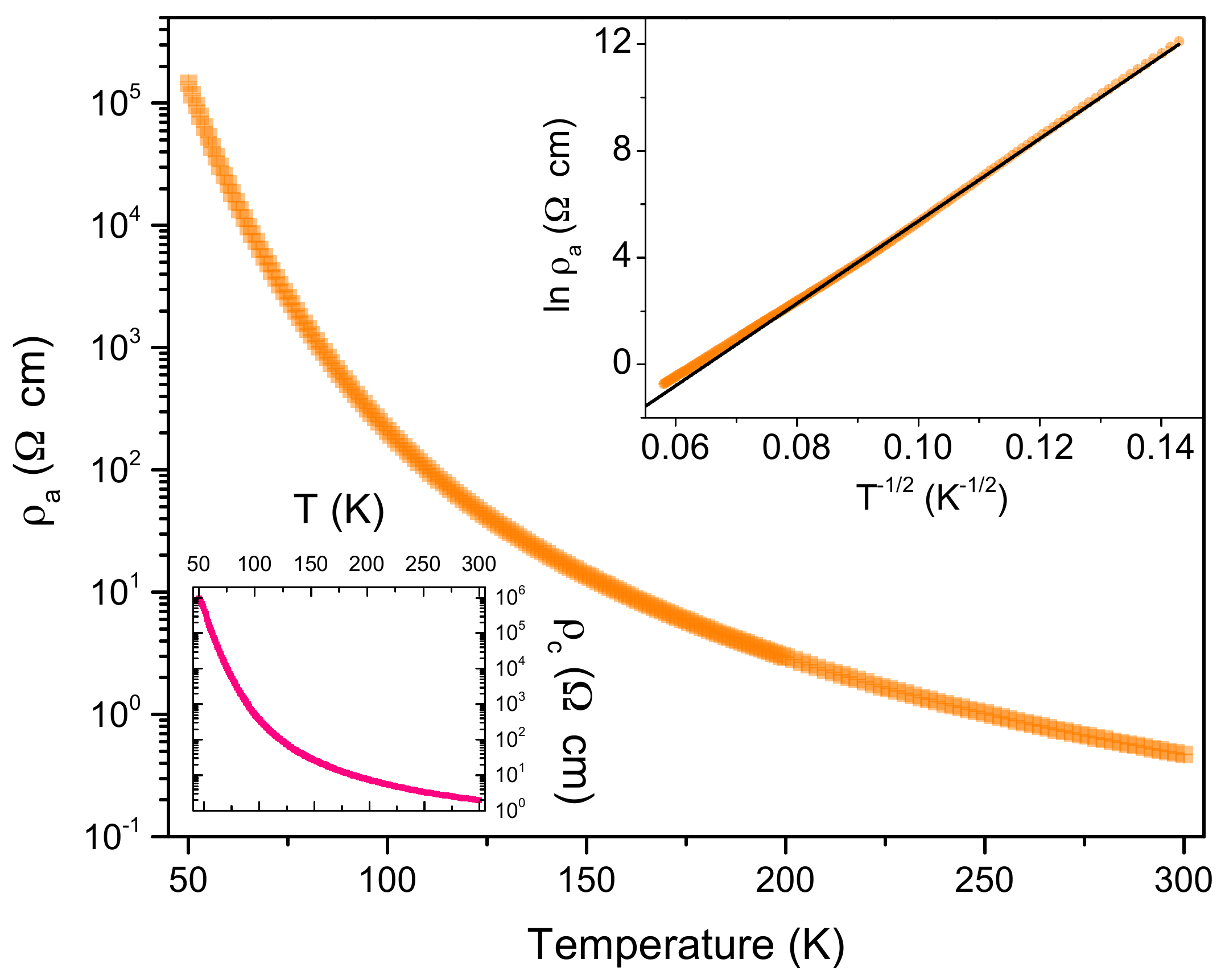}
\caption{Temperature dependence of resistivity measured along the $a$ and $c$-axes of (Ca$_{0.89}$Sr$_{0.11}$)$_5$Ir$_3$O$_{12}$.  The main panel shows $\rho_a$ along the $a$-axis and the lower left inset plots $\rho_c$ along the $c$-axis. Upper inset shows plot of $ln\rho_a$ versus $T^{-\frac{1}{2}}$ and the corresponding VRH fit of resistivity with $D=1$ and $T_0=23800$K.}
\end{figure}

Figure $10$ shows charge transport measurements along the $a$ and $c$ directions of a (Ca$_{0.89}$Sr$_{0.11}$)$_5$Ir$_3$O$_{12}$ crystal. Insulating behavior was observed along both the $a$ and $c$ axes across all temperature ranges probed.  Transport along the $a$-axis is shown in Fig. 10's main panel while $c$-axis transport data are plotted in the lower left inset. We note here that due to the constraints of the samples' dimensions, $c$-axis transport data could only be acquired via a two-probe measurement whereas $a$-axis measurements used a conventional four probe measurement.  Contact resistance in the former case should be a small percentage of the resistance values at low temperature, and comparison of the $a$ and $c$ axis low temperature transport data in this regime reveals minimal anisotropy.

$a$-axis charge transport data were best fit with a VRH model with $D=1$ as shown in Fig. 10's upper right inset. This naively agrees well with the quasi $1$-D nature of this compound; however the absence of appreciable anisotropy in charge transport instead suggests that the Efros-Shklovskii (E.S.) form of VRH\cite{0022-3719-8-4-003} is the more likely transport mechanism.  A similar analysis of the $c$-axis transport data also best fits to the ES form of tunneling transport, $\rho(T) \propto e^{(\frac{T_0}{T})^{1/2}}$.  Magnetization data are plotted in Fig. 11 where an antiferromagnetic transition at T$_{AF}=9(1)$ K appears via the onset of irreversibility between FC and ZFC data (Fig. 11 (a)) with the field applied either within the $ab$-plane or along the $c$-axis.  CW fits were performed to the data above this transition using the form $\chi = \frac{C}{T-\Theta}+\chi_0$ where $\chi_0$ encodes the weak core level diamagnetism, any temperature independent Van Vleck contribution, and the background contribution from the sample mount.

\begin{figure}
\includegraphics[scale=.38]{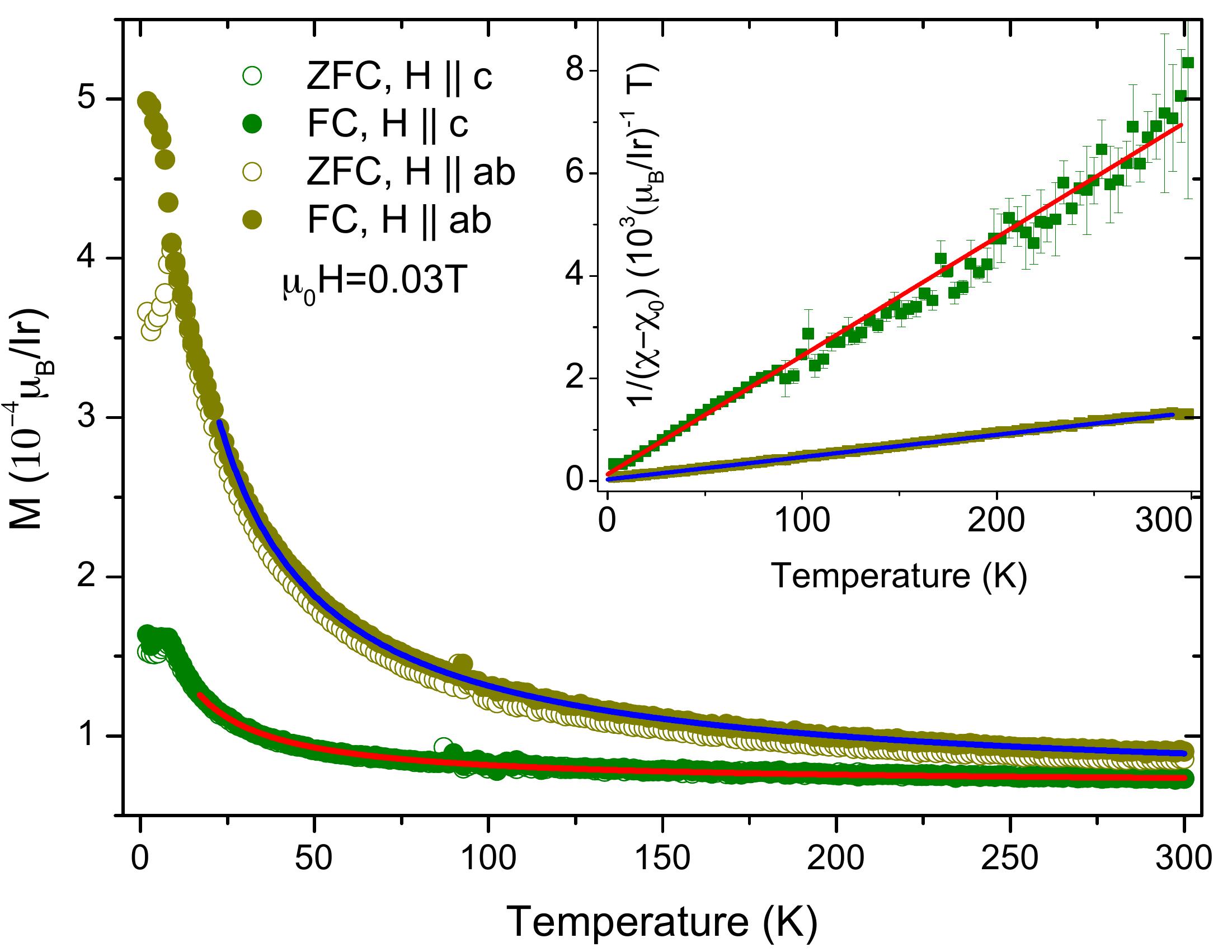}
\caption{Temperature dependence of the magnetization of (Ca$_{0.89}$Sr$_{0.11}$)$_5$Ir$_3$O$_{12}$ collected under FC (closed symbols) and ZFC (open symbols) with $\mu_0 H=0.03$ T applied within the $ab$-plane and along the $c$-axis.  Inset shows the corresponding inverse susceptibilities with temperature independent background and Van Vleck terms removed, $1/(\chi - \chi_0)$. CW fits to the data are plotted as solid lines.}
\end{figure}

The upper inset in Fig. 11 (a) shows the results of CW fits along both directions which yielded $\Theta_{CW}=-7.7(1)$ and $-5.8(2)$ K for the magnetic field oriented within the $ab$-plane and along the $c$-axis respectively. The temperature independent $\chi_0$ terms were identical within error for both field orientations and was comprised of intrinsic and extrinsic background contributions where $\chi_0 = \chi_{intr} + \chi_{bckg}$.  The $\chi_{bckg}$ was measured independently and the $\chi_{intr}$ contributions were $\chi_{intr} = 0.00234(1) \mu_B$/Ir and $\chi_{intr} = 0.00224(1) \mu_B$/Ir for the $c$-axis and $ab$-plane field orientations respectively.  Curiously, the local moments extracted from the $C$ values were substantially different for the two field orientations.  In accounting for the mixed valence of Ir-cations inherent to the Ca$_5$Ir$_3$O$_{12}$ structure (ie. 1/3 of the Ir cations are Ir$^{4+}$ and 2/3 are Ir$^{5+}$), the local moments from CW fits should arise from the Ir$^{4+}$ sites alone. Correcting for this gives $\mu_{eff}=1.76\pm0.02$  $\mu_B$ for Ir$^{4+}$ moments in measurements with the field applied within the $ab$-plane and $\mu_{eff}=0.76\pm0.02$  $\mu_B$ for measurements with the field applied along the $c$-axis.  This implies a substantial orbital component to the moments where an anisotropic $g$-factor suppresses the apparent local moment along the chain direction.

\section{Discussion}

Figure 12 summarizes the structural phase evolution of crystals yielded as the nominal starting ratio of 2(Sr(1-x):Ca(x))CO$_3$:IrO$_2$:6SrCl$_2$ was varied.  Symbols denote the typical phase content of individual crystals harvested within a batch, and overlaps between structure types denote regions where individual crystals contained of mixture of two phases.   The relative molar fractions of crystals in this mixed phase region are indicated by the phase percentage axis.  The $n=1$ R.P. phase remains stable until $x_{nominal}\approx11 \%$, where increased Ca-substitution into (Sr$_{1-x}$Ca$_x$)$_2$IrO$_4$ enhances the weak ferromagnetism of the $x_{nominal}=0$ parent system.  Beyond this starting Ca-composition, a miscibility gap opens and the dominant phase yielded in crystal growth was the Sr$_3$CaIr$_2$O$_9$ state, yet crystals still included intergrowths of $n=1$ R.P. phase for concentrations $x_{nominal}\leq0.5$.  Phase pure Sr$_3$CaIr$_2$O$_9$ crystals nucleated in the composition range $0.5\leq x_{nominal}\leq0.7$ before switching to the quasi one-dimensional Ca$_{5}$Ir$_3$O$_{12}$ structure beyond $x_{nominal}=0.9$.

Within the $n=1$ R.P. phase, Ca-substitution into Sr$_2$IrO$_4$ serves as only a small perturbation to the lattice where the unit cell volume contracts by only $0.6 \%$ under approximately $10 \%$ Ca-substitution.  When compared to previous hydrostatic pressure studies that affect a much larger volume reduction, \cite{Haskel2012}  the relative reduction in the low temperature resistivity of Ca-substituted samples is substantially greater than externally applied pressure.  This verifies that the enhanced conductivity is defect-driven, where the added A-site disorder generates states within the gap.  The low temperature flattening in the resistivity further suggest a percolative network formed by defect states; however a crossover to metallic transport was not observed up to the solubility limit of Ca within the Sr$_2$IrO$_4$ lattice.  The persistence of magnetic order up to this limit further supports this picture of a small volume of defects dominating transport data, where previous hydrostatic pressure studies resolved the collapse of AF order at pressures far below the collapse of the charge gap in this compound.\cite{Haskel2012}

The enhancement of the weak ferromagnetic moment observed in Ca-substituted Sr$_2$IrO$_4$ cannot be trivially accounted for by assuming a fixed ordered moment size locked to the in-plane octahedral rotation.  Rather, a change in the relative coupling between degree of spin canting and octahedral rotation likely occurs.  This enhancement in the canting of moments relative to the IrO$_6$ octahedral rotations can potentially be driven via an enhanced tetragonal distortion of the ligand field at the Ir$^{4+}$ sites.\cite{PhysRevLett.102.017205}  As expected with the substitution of the small Ca-cation within the rock-salt layers of the $n=1$ R.P. structure, our PXRD refinements reveal that a slight elongation ($\approx 2\%$) of the octahedra results near the $x=0.11$ limit of Ca-substitution.  However, this enhanced elongation should naively reduce the canting angle of the moments relative to the octahedral rotation, highlighting the need for detailed neutron diffraction data sensitive to the ordered moment size and oxygen positions to fully resolve the mechanism of the apparent moment enhancement.

\begin{figure}
\includegraphics[scale=.375]{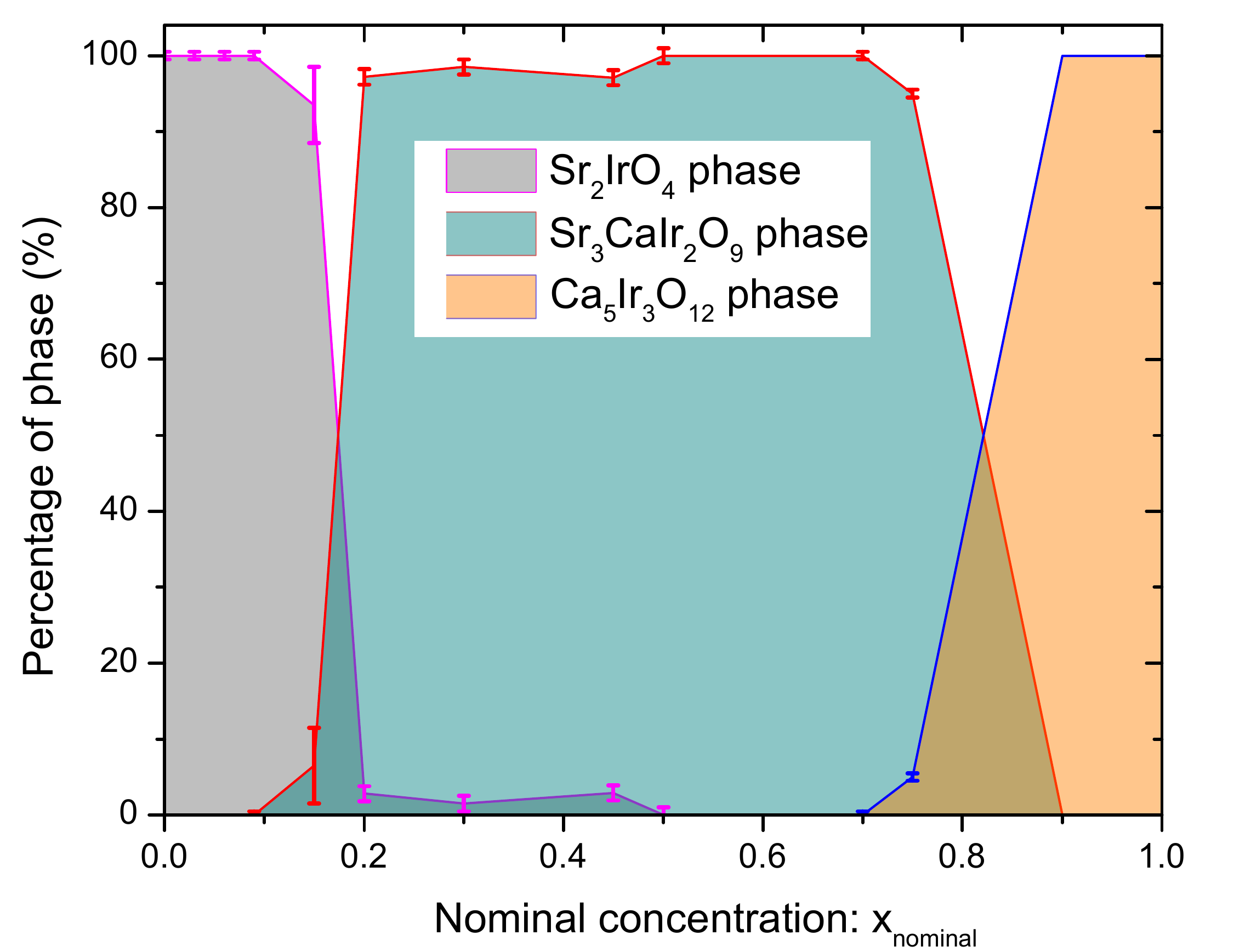}
\caption{ Phase diagram generated based on the growth sequence and procedures detailed in Section II.  Predominant structure types of (Sr$_{1-x}$Ca$_x$)$_{2-y}$IrO$_{4+z}$ that crystallize in each region of the phase diagram are shaded as grey (Sr$_2$IrO$_4$), teal (Sr$_3$CaIr$_2$O$_9$), and peach (Ca$_5$Ir$_3$O$_{12}$).  Regions of overlap denote concentrations where mixed phase crystals were obtained with the relative phase fractions denoted on the phase percentage axis.  We note here that the bottom axis denotes the nominal, starting concentration ($x_{nominal}$) of Ca in the synthesis run.}
\end{figure}

Data collected on crystals grown within the Sr$_3$CaIr$_2$O$_9$ structure agree with an earlier powder study of this material. \cite{Wallace2015}   No relaxation of the structure into a higher symmetry space group was observed as the system was heated to 473 K; however our PXRD measurements have poor sensitivity to the oxygen site positions and occupancies. Minimal anisotropy was observed in both transport and magnetization measurements, and our magnetization data are consistent with the picture of Sr$_3$CaIr$_2$O$_9$ possessing a non-magnetic $J=0$ ground state.  The weak moments observed in magnetization measurements are consistent with a trace amount of oxygen vacancies generating small fractions of Ir$^{4+}$ sites and $J=1/2$ moments.  This was indicated by the small and variable scaling factors $\eta \approx0.02$ needed to account for Brillouin function fits of the isothermal magnetization data at $T=5$K in different crystals.

For crystals that nucleated in the one-dimensional Ca$_5$Ir$_3$O$_{12}$ structure at the Ca-rich end of the phase diagram, the substitution of Sr into the lattice seems to have a substantial influence on the physical properties of the material.  Specifically, for $11 \%$ Sr-substitution onto the Ca-sites, a number of electronic properties observed differ from an earlier report studying the unalloyed $x=1$ system. \cite{Cao2007PRB}  No thermally-driven metal to insulator was observed in $ab$-plane transport data in our Sr-substituted crystals, where instead the entirety of resistivity data was modeled via E.S. VRH.  Susceptibility data do not reveal any broad maxima consistent with a one-dimensional quantum antiferromagnet. \cite{PhysRev.135.A640}  Rather CW fits of Sr-alloyed crystals reveal $\Theta_{CW}$ values in accordance with the onset temperature of AF order at T$_{AF}=9(1)$K, in contrast to the large $\Theta_{CW}\approx 280$ K and high frustration parameter reported  for the $x=1$ system.  While the anisotropy in the low field magnetization data of the $x=0.89$, $11 \%$ Sr-substituted, system matches that of the $x=1$ parent compound, the anisotropy observed in the apparent $\mu_{eff}$ values of the $x=0.89$ system under different field orientations suggests a highly anisotropic $g$-factor for this system.  Future work exploring this via paramagnetic spin resonance techniques directly sensitive to the $g$-tensor are required to fully explore the origin of this effect.

\section{Conclusions}

In summary, the structural and electronic phase evolution of (Sr$_{1-x}$Ca$_x$)$_{2-y}$IrO$_{4+z}$ crystals was explored.  Ca substitution into the spin-orbit Mott insulator (Sr$_{1-x}$Ca$_x$)$_2$IrO$_4$ was possible for $x_{nominal}\leq0.11$ before a miscibility gap opens and the Sr$_{3}$CaIr$_2$O$_9$ phase intermixes. Prior to this, the spin-orbit Mott insulating state survives the relatively weak perturbation to the structure driven via Ca impurities within the lattice.  An anomalous enhancement of the net ferromagnetic moment results with Ca substitution, suggestive of a decoupling between the in-plane canted antiferromagnetism and the in-plane IrO$_6$ octahedral rotation. As the nominal Ca substitution level is increased to $x_{nominal}=0.5$, phase pure Sr$_3$CaIr$_2$O$_9$ nucleates and manifests a nonmagnetic $J=0$ insulating ground state consistent with recent powder studies.  At higher nominal Ca concentrations $x_{nominal}>0.7$, the one dimensional chain structure (Sr$_{1-x}$Ca$_{x}$)$_5$Ir$_3$O$_{12}$ stabilizes. For partially alloyed crystals in this system ($x=0.89$), the low temperature insulating ground state reveals an antiferromagnetic phase transition below $T =9$ K with an anomalous high temperature anisotropy in the CW susceptibility not present in the $x=1$ system.  Future studies exploring the electronic phase evolution close to the $x=1$ end point of this system are a promising future direction in understanding the origin of this response.

\acknowledgments{
This work was supported by NSF Award No. DMR-1505549.  X.C. gratefully acknowledges the helpful discussions with Guang Wu and the 11-BM beam line staff for assistance with experiments.  Use of the Advanced Photon Source at Argonne National Laboratory was supported by the U. S. Department of Energy, Office of Science, Office of Basic Energy Sciences, under Contract No. DE-AC02-06CH11357.  The MRL Shared Experimental Facilities are supported by the MRSEC Program of the NSF under Award No. DMR 1121053; a member of the NSF-funded Materials Research Facilities Network.}

\bibliography{CaSr214}

\begin{thebibliography}{67}
\bibitem[{Sup()}]{Supplemental}
\bibinfo{note}{We define the measured direction as (1, 1, 1), however due to
  the low symmetry, there are other possible orientations: (-3, 1, 1); (3, 1,
  -1); (0, 0, 2); (4, 0, -2); (-1, 1, -1). Nevertheless no appreciable
  difference is observed when measurements along orthogonal directions of the
  crystal were performed.}
\end{thebibliography}
%
%
\end{document}